# Mapping of ion beam induced current changes in FinFETs

C. D. Weis[1,2], A. Schuh[1,2], A. Batra[1], A. Persaud[1], I. W. Rangelow[2], J. Bokor[3,4], C. C. Lo[3], S. Cabrini[4], D. Olynick[4], S. Duhey[4], and T. Schenkel[1]*

[1]Lawrence Berkeley National Laboratory, 1 Cyclotron Road, Berkeley, CA 94720, USA

[2]Technical University Ilmenau, D-98684 Ilmenau, Germany

[3]Department of Electrical Engineering and Computer Science, University of California, Berkeley, CA 94720, USA

[4]The Molecular Foundry, Lawrence Berkeley National Laboratory, Berkeley, CA 94720, USA

**Abstract:**

We report on progress in ion placement into silicon devices with scanning probe alignment. The device is imaged with a scanning force microscope (SFM) and an aligned argon beam (20 keV, 36 keV) is scanned over the transistor surface. Holes in the lever of the SFM tip collimate the argon beam to sizes of 1.6 μm and 100 nm in diameter. Ion impacts upset the channel current due to formation of positive charges in the oxide areas. The induced changes in the source-drain current are recorded in dependence of the ion beam position in respect to the FinFET. Maps of local areas responding to the ion beam are obtained.

---

[1]* Corresponding Author: Thomas Schenkel, Email: T_Schenkel@LBL.gov, Phone: +1-510-486-6674, FAX: +1-510-486-5105

1. Introduction

Kane's proposal of a quantum computer based on donor spin qubits in silicon [1] requires the precise positioning of single dopant atoms in devices integrated with control gates and readout structures. Further, controlled doping of conduction channels in conventional CMOS technology is of interest because device structures are scaled down continuously and homogeneous doping profiles can no longer be assumed and cause fluctuations in device performance [2-5]. We recently demonstrated a spin to charge conversion mechanism with potential for scaling to single spin readout using field effect transistors (FET) [6]. In addition we demonstrated a technique for single ion doping of these FETs by monitoring changes in the source-drain currents at room temperature [7]. The whole transistor was exposed to an ion beam and the position of impinging ions into the channel of the FET was defined by a hole in the gate electrode for the detection of single ions. Only ions within this region reached the 2DEG channel and could be detected. Our goal is to place single ions at precise positions by using an SFM with an integrated ion beam (Fig. 1) [8, 9]. This will allow us to scan our devices with no ions hitting the surface, and then align the ion beam to desired implantation regions and implant ions deliberately instead of relying on aperture holes in the gate electrode. The ion beam is collimated by holes in a pre-collimator and the cantilever of the SFM tip. The next steps towards ion positioning are scanning this ion beam over the active area of transistors and mapping those regions. This approach is common in ion beam induced charge (IBIC) mapping. IBIC can be used as a tool to investigate the uniformity of charge transport in bulk semiconductors; and the influence of extended features such as grain boundaries, precipitates, and twins on charge transport [10-12]. Here we describe our first aligned ion implantation onto a surface feature that acts as a local detector using our SFM tip. This is a step towards precise positioning of implanted ions and will allow calibration of tip beam distances in the future.

2. Experimental setup

FinFETs were fabricated in 250 nm thick SOI (200 nm 28Si epi layer on 50 nm natural silicon on oxide). 100 nm low temperature oxide was deposited to serve as the hard mask for the silicon etch. Source/drain pads and channel (fin) regions were defined by electron beam lithography. Channel widths ranged from 30-250 nm. After the silicon etch, 10 nm gate oxide was grown on the sidewalls. In situ phosphorus-doped polycrystalline silicon was deposited as gate electrodes with a thickness of 140 nm. Gates of 280 nm in length were patterned by e-beam lithography and poly silicon dry etching. Self-aligned arsenic implants (25 keV, $2 \times 10^{15}$ / cm²) were used to form source/drain regions. Low-temperature chemical-vapor deposited silicon dioxide (LTO) was used as an interlayer dielectric layer with a thickness of 200 nm. Contact regions were etched and tungsten was used for metal contacts. Devices were annealed in forming gas ($N_2/H_2$-90%/10%) at 400°C for 20 min to passivate defects at the $Si/SiO_2$ interface and to improve the metal-semiconductor contact quality. Fig. 2a shows the FinFET before LTO film deposition and Fig. 2b shows an *in-situ* SFM image after LTO deposition. A dual beam focused ion beam (FIB), consisting of a 30 keV $Ga^+$ ion beam and a 5 keV electron beam with $XeF_2$ etch gas supply, was used to remove parts of the LTO layer and gate electrode. After FIB processing, the devices were annealed in forming gas ($N_2/H_2$-90%/10%) at 400°C for 20 min. Electrical device performance was recovered after the anneals (see Fig. 3). Note that the data after FIB processing is multiplied by a factor of 50 and that the pristine IV-curve is not displayed. Gate leakage currents were below 200 pA. FinFETs were then mounted in our ion implantation setup with aligned SFM. Argon ions were produced in an electron cyclotron resonance (ECR) source [13]. $Ar^{3+}$ (36 keV) and $Ar^{2+}$ (20 keV) ions were extracted and selected with a 90° analyzing magnet and focused onto the sample. Pulsing of the ion beam was achieved by applying a voltage onto a deflector plate parallel to the ion beam trajectory. The devices were operated at a gate bias $V_g$ of 1.0 V, a drain bias $V_d$ of 0.8 V, and with the source grounded, resulting in drain currents of approximately 10 µA. The drain current signal was recorded after amplification in a current amplifier (Stanford Research 570). The mounted device was scanned with the SFM and the tip was moved over the active area of the FinFET. An offset in x- and y-position was applied so that one of the drilled holes in the SFM cantilever aligned with the active area. The distances and directions were obtained from scanning microscope images (SEM), which showed the SFM cantilever and the holes next to the imaging tip. The tip was then moved along each spot of a square point grid. At each spot, the drain current was measured over time, and the argon ion beam was turned on after a dwell time in which no ions hit the sample. SFM tip position and drain currents were recorded and ion beam exposures were controlled with a Labview program. The total absolute change in the drain current within each interval was plotted vs. x- and y-position of the imaging tip to generate the IBIC maps.

3. Results

The data of three consecutive exposure spots is displayed in Figure 4. Argon ions (36 keV, collimated with the 1.6 μm hole) hit the FinFET during intervals, which are indicated with arrows. The y-position is kept constant and the x-position is changed by 1 μm between spots. Intervals a) and b) show an increase in drain current. No such increase, just the fluctuation of the channel current, can be seen in interval c). In cases a) and b), the tip was in a position in which ions hit areas that resulted in an increase in drain current. As previously demonstrated, ion hits can form positively charged traps in oxides which alter transistor currents [14]. Those positive oxide charge traps increase the effective gate voltage. The three data sets were obtained when the beam was initially aligned to the gated source-drain region and moved away from that active region. The full set of data for that scan is shown in Fig. 5. Data from 21 x 21 spots were recorded in a 20 x 20 μm² area. The exposure time was 5 seconds for each spot. On average, ~600 ions/sec/μm² hit the sample and the peaks that correspond to changes in the source-drain current are clearly visible. The beam spot is larger than the distance between consecutive spot locations. Therefore, nearest neighbor spots contain responses from the same area. The beam diameter is fairly large compared to the dimensions of the source-drain channel. In order to obtain more accurate maps of the channel region, we collimated the beam with a different hole in the SFM lever which was ~100 nm in diameter. Fig. 6 shows one of the IBIC maps obtained with argon ions (20 keV). The exposure time was set to 30 sec per spot and the beam fluence was ~1500 ions/sec/μm². The distance between spots was 200 nm and 16 x 16 spots were exposed within a 3 x 3 μm² area. The signal ratio of active area to surrounding area is smaller than in Fig. 5. The fluence of the 20 keV ions was 2.5 times higher than the fluence of the 36 keV ions, but the energy of the 20 keV ions was 44% less. The expected responses would be similar under those conditions. The smaller response is explained by the difference in the size of the holes in the cantilever. In the case of the smaller hole, only parts of the total active area are exposed, whereas beam spots collimated with the larger hole exposed most of the entire active area at once. In both circumstances, four beam positions resulted in a channel current response. The large hole beam-spots were overlapping whereas the small hole beam-spots were separated. The active area can be estimated to be around 400-600 nm in side length. This is larger than we would have expected from the fin dimension of 100 x 280 nm². No single ion hits were observed and relatively high ion fluences were needed to observe changes in the channel current. One possible explanation for the larger result is that the LTO and gate material on top of the fin were not removed completely. Impinging ions may have created positively charged defects in the LTO layer instead of the gate oxide as previously observed [7].

4. Discussion and outlook

This paper shows aligned ion implantation onto a surface feature that acts as a local detector. Thus far, patterns were generated on samples without *in-situ* ion detection [8, 9, 15], and impinging ions on the surface were detected while the whole sample was exposed [7]. Here, the ion beam was pre-collimated, the desired implantation region was scanned with the SFM, the tip was aligned to the source-drain region and the ion beam was scanned over a small area to map the responsive areas of our devices. This is an important step towards our goal of precision ion implantation and will act as an *in-situ* calibration method for our tips in the future. The resolution of the maps will be increased by collimating the beam further down. Calibration will improve the positioning of the tip and alignment of the ion beam to selected implant locations. We are in the process of improving the LTO and gate oxide removal in order to combine the detection of single ions with the IBIC mapping. This will enable us to study the effects on device performance, e.g. change in the threshold voltage, caused by the number and position of donor atoms. Since the devices survive activation anneals, consecutive implantations are possible and will be monitored and correlated to changes in device behavior.

**Acknowledgments:**

We thank the staff of the Microlab at UC Berkeley and the staff of the Molecular Foundry and the National Center for Electron Microscopy at LBNL for their support. This work was supported by NSA under contract number MOD 713106A, the National Science Foundation through NIRT Grant No. CCF-0404208, and by the Director, Office of Science, of the Department of Energy under Contract No. DE-AC02-05CH11231.

**Figure Captions:**

Figure 1: Schematic of the IBIC mapping procedure.

Figure 2: SEM image of a FinFet before LTO deposition (top). *In-situ* SFM image after LTO deposition (bottom).

Figure 3: Room temperature I-V-curves of a FinFET after FIB processing and after annealing in forming gas.

Figure 4: Example of absolute change of the source-drain current during IBIC map acquisition depending on the x- (blue) and y-positions (green) of the AFM tip.

Figure 5: IBIC map obtained with an argon beam collimated by a 1.6 μm hole in the SFM lever.

Figure 6: IBIC response recorded with an argon beam collimated by a 100 nm hole in the SFM lever.

**Figures:**

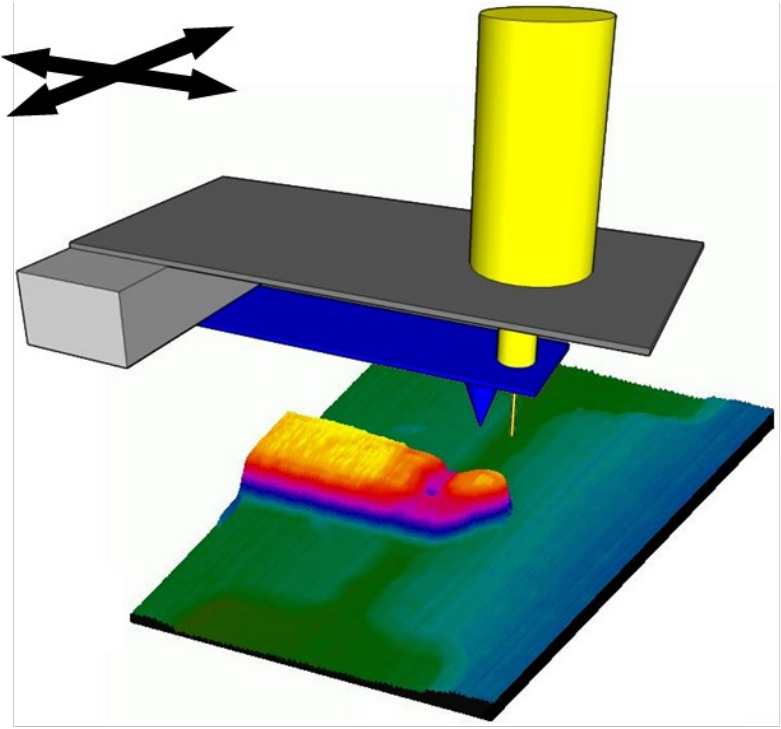

Figure 1

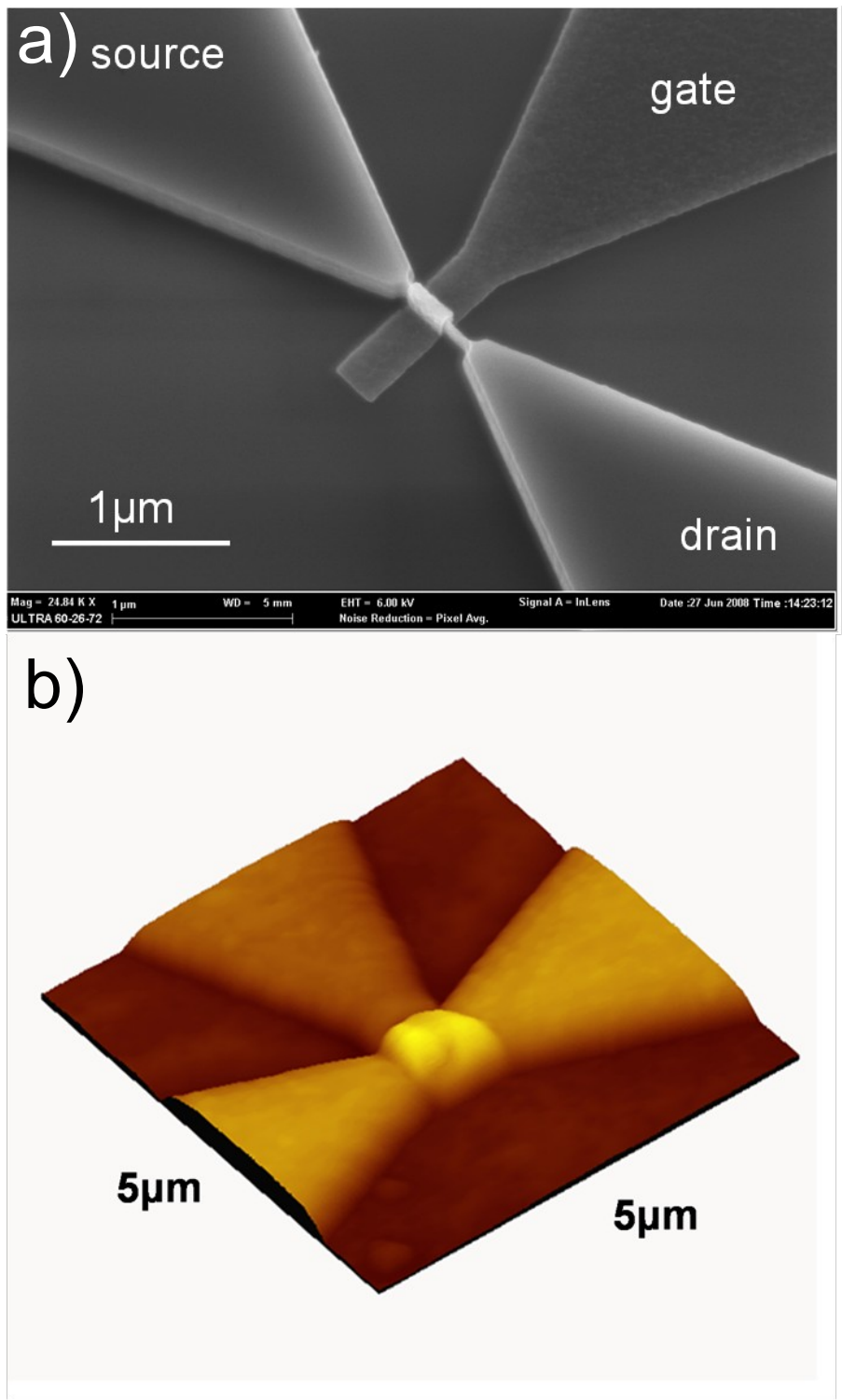

Figure 2

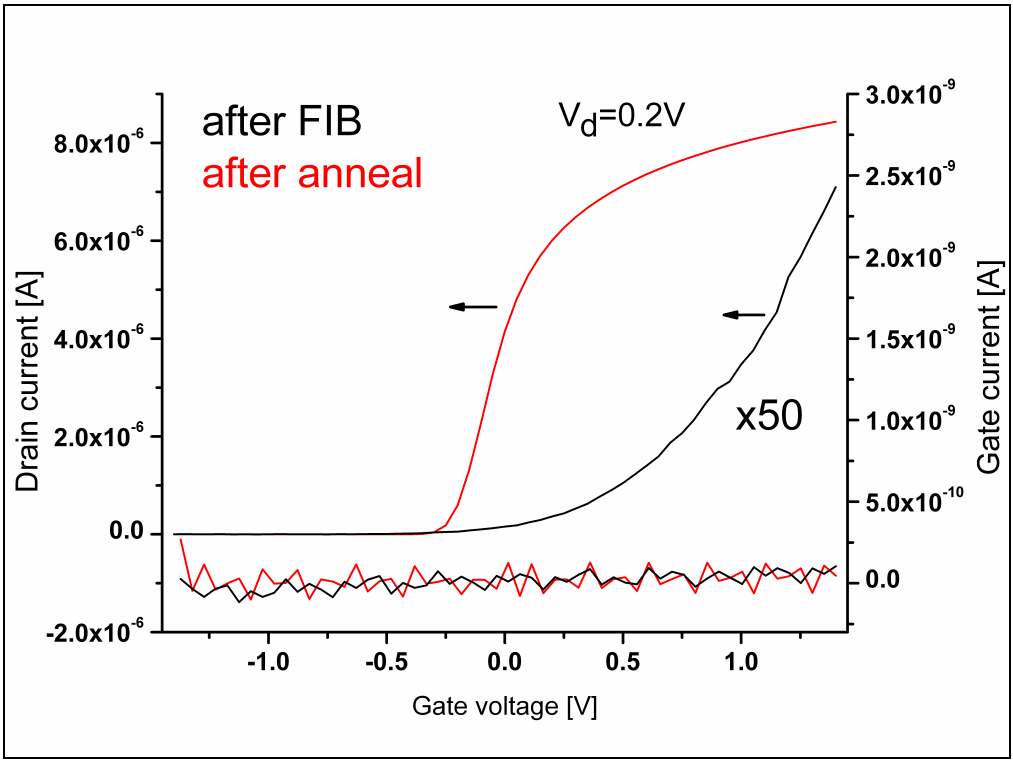

Figure 3

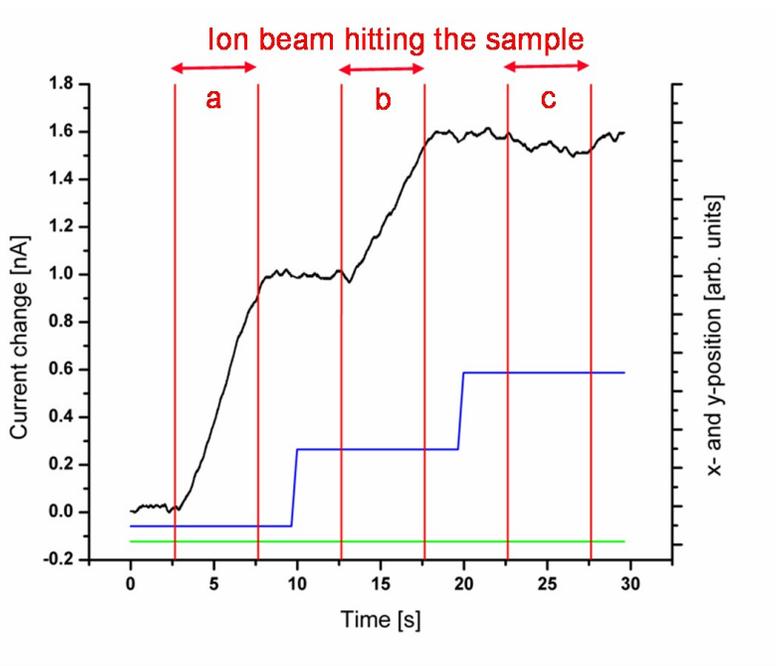

Figure 4

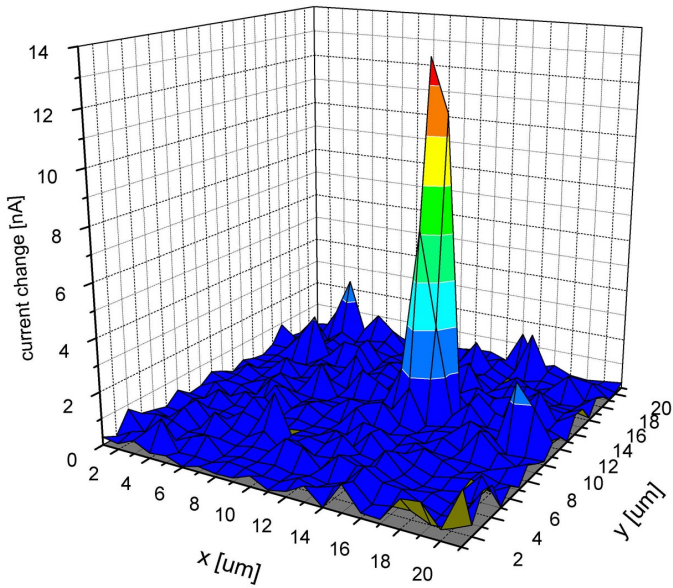

Figure 5

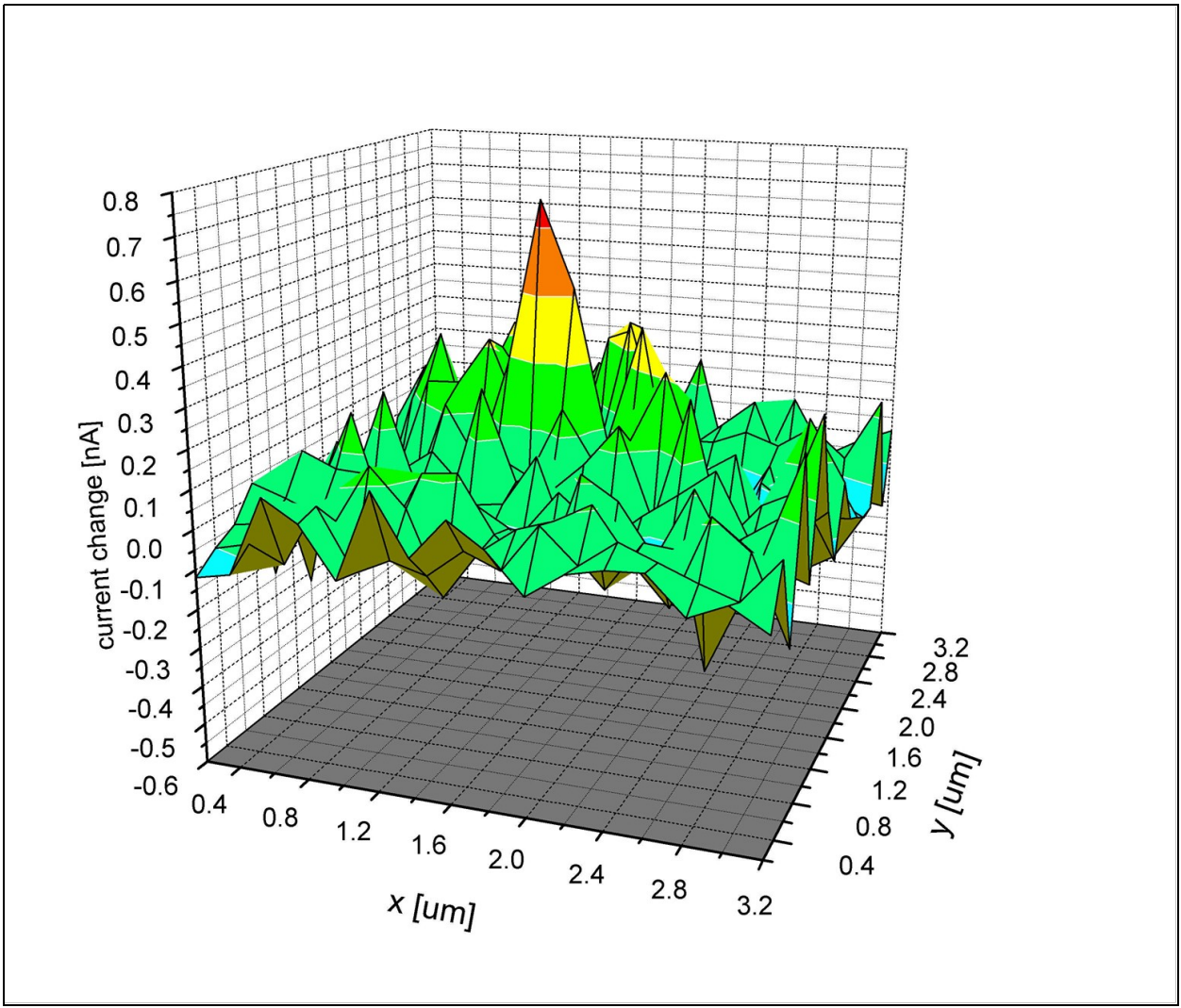

Figure 6